\documentstyle[11pt,multicol]{article}
\input epsf
\def\paper#1#2#3#4#5{#1, #2 {\bf #3} (#5) \rm #4}

\def\be{\begin{equation}}
\def\ee{\end{equation}}
\def\bsigma{\mbox{\boldmath $\sigma$}}
\def\uhp{{\sc uhp}}
\def\fbc{{\rm f\sc bc}}
\def\Fbc{{\sc Fbc}}

\begin{document}
\title{
  \begin{center}
    {\huge { Evidence for a topological transition in 
         nematic-to-isotropic
         phase transition in two dimensions}}
  \end{center}
}
\author{{\normalsize A.I. Fari$\tilde{\rm n}$as Sanchez$^{a,b}$, 
  R. Paredes V.$^{a}$ and B. Berche$^{b}$}\\[0cm]
  {\small\it $^{a}$ Centro de F\'\i sica, Instituto Venezolano de Investigaciones 
    Cient\'\i ficas,}\hfill\ \\[-1mm]
  {\small\it Apartado 21827, Caracas 1020A, Venezuela}\hfill\ \\[0mm]
  {\small\it $^{b}$ Laboratoire de Physique des Mat\'eriaux, 
    Universit\'e Henri Poincar\'e,  Nancy 1,}\hfill\ \\[-1mm]
  {\small\it F-54506 Vand\oe uvre les Nancy Cedex, France}\hfill\ \\[0.3cm]}
\maketitle
\vspace{-1cm}
\begin{abstract}
  {\small The nematic-to-isotropic orientational phase transition, or
    equivalently the $RP^2$ model, is considered
    in two dimensions and the question of the nature of the phase transition
    is addressed. Using powerful conformal techniques adapted to the
    investigation of critical properties of two-dimensional scale-invariant 
    systems, we report strong evidences for a transition governed
    by topological defects analogous to the Berezinskii-Kosterlitz-Thouless
    transition in two-dimensional $XY$ model.
  }
\end{abstract}
\begin{multicols}{2}
Liquid crystals may be seen as constituted of molecules essentially represented
by long rigid rods. From maximization of entropy at high temperatures, all 
the molecule orientations are equally probable, independently of the 
neighbouring molecule directions and the system exists in an isotropic phase. 
At low temperatures a preferential orientation is more favourable in order
to minimize interaction terms, and an ordered structure emerges. When order
occurs along one space dimension only, the system is said to be nematic.
Still at lower temperatures, other ordered phases can appear, e.g. smectic
phases.

In a lattice model, each molecule may be represented by a unit vector 
$\bsigma_w$ at site $w$ of an hyper-cubic lattice $\Lambda$ of linear extent 
$L$. The $\bsigma$'s live in a three-dimensional space attached to each lattice
site. In the nematic phase, the preferential direction defines a unit
vector, ${\bf n}$, called the director, and one can measure the deviation of
molecule $\bsigma_w$ with respect to the director by the scalar product
$\bsigma_w\cdot{\bf n}=\cos\theta_w$. Due to the local $Z_2$ symmetry (the
rods are not oriented), one cannot distinguish between opposite directions 
$\theta_w$ and $\theta_w+\pi$, and $\cos\theta_w$ vanishes on average while
$\cos^2\theta_w$ does not.
In the disordered phase on the other 
hand, the angles are measured with respect to any arbitrary direction, and
the thermal average of course leads to $\langle\cos\theta_w\rangle=0$ and
$\langle\cos^2\theta_w\rangle=\frac 13$, so that $\langle\cos^2\theta_w\rangle
-\frac 13$ represents 
a convenient order parameter. In the literature on liquid crystals, one
usually defines the local order parameter by the second Legendre polynomial,
\be m(w)=\langle P_2(\bsigma_w\cdot{\bf n})\rangle =\langle P_2(\cos\theta_w)
\rangle.\ee
This definition suggests to consider the following Hamiltonian to describe
the nematic transition,
\be
-\frac{H}{k_BT}=\frac{ J }{k_BT}\sum_w\sum_\mu P_2(\bsigma_w
\cdot\bsigma_{w+\mu}),\label{HamLL}\ee
where $\mu$ stands for the unit basis vectors of the lattice, $\bsigma_w
\cdot\bsigma_{w+\mu}=\cos(\theta_w-\theta_{w+\mu})$ is the scalar product
between neighbouring vectors distant from one lattice spacing, and the
interaction term $- J  P_2(\bsigma_w\cdot\bsigma_{w+\mu})$ is reminiscent
from a dipole-dipole interaction. This Hamiltonian was introduced
by Lebwohl and Lasher~\cite{LebwohlLasher73} as a lattice version of the
mean field theory of Maier and Saupe~\cite{MaierSaupe59}, and its success
came from its ability to reproduce the weak first order phase transition
observed experimentally in the three dimensional nematic 
transition~\cite{ZhangMouritsenZuckermann92}. In a more abstract context,
this Hamiltonian is known as the $RP^2$ model, since at each lattice site
is attached the manifold of directions in 3-dimensional space, also called the
real projective space in 3 dimensions~\cite{KunzZumbach92}.

Like the non linear $\sigma$-model, this model possesses generically the 
symmetry group $O(n)$ which is non abelian for $n\ge 3$, and specifically 
the Lebwohl-Lasher or $RP^2$ model has a $O(3)$ symmetry. 
The question, as it was outstandingly 
formulated by Kunz and Zumbach~\cite{KunzZumbach92}, of the nature of 
the transition in 
the $RP^2$ model in two dimensions is still incompletely solved. 
The existence of a
phase transition (at finite temperature) in two dimensional systems seems 
connected to the abelian nature of the underlying symmetry group, as 
both Ising and $XY$ models are famous examples, unlike the Heisenberg model. 
On the other hand, according to the Hohenberg-Mermin-Wagner 
theorem~\cite{MerminWagner66,Hohenberg67}, models possessing a continuous
symmetry group cannot exhibit any finite macroscopic magnetization with
no magnetic field applied in dimensions 1 or 2 (we intentionally use the 
familiar terminology of  magnetic systems). The two-dimensional $XY$ model
is the most famous example and it exhibits a non conventional 
transition~\cite{KosterlitzThouless78,Nelson83,ItzyksonDrouffe89}.
In spite of the absence of long-range order (LRO) at low temperatures,
the spin-spin correlation function decays algebraically
with an exponent which increases monotonically with 
temperature~\cite{Rice65,Wegner67,Sarma72} up to a temperature called after
Berezinskii, Kosterlitz and 
Thouless~\cite{Berezinskii71,KosterlitzThouless73,Kosterlitz74}. In this 
critical phase, macroscopic ordering is prevented by collective excitations, namely 
spin waves 
which nevertheless do not exclude a coherent 
orientation of spins at a smaller length scale. Together with the spin 
waves, localized excitations appear with increasing temperature. 
These are topological defects associated in pairs, like pairs of opposite 
charges in the low temperature phase of a two-dimensional Coulomb gas, 
and they perturb the spin field only locally. This is the usual meaning of 
the term quasi-long-range order (QLRO).
The Berezinskii-Kosterlitz-Thouless transition is governed by unbinding of
these defects which completely suppresses any type of long-range order at
high temperature, hence the correlation functions decay exponentially like
in an ordinary paramagnetic phase. This is the standard scenario of a
topological transition.
The puzzle becomes confused when one notices that $XY$ or Heisenberg models
in three dimensions display conventional continuous transitions and the 
$RP^2$ model exhibits a first order transition while renormalization group
treatment of non-linear $\sigma$ model predicts the absence of any transition
at non-zero temperature 
in $2d$ (asymptotic freedom in the context of lattice gauge theories~\cite{Kogut79}) 
and a continuous one in $3d$ for all the three models.
We also have to mention that the $RP^1$ model (the same as given in 
equation~(\ref{HamLL}), but with two-component vectors $\bsigma_w$) 
exactly coincides with the $XY$ model.
The question of a possible topological transition in the two-dimensional
(non abelian) Lebwohl-Lasher model is thus particularly attracting and was already addressed in previous 
studies~\cite{DuaneGreen81,SolomonStavansDomany82,ChiccoliPasiniZannoni88,KunzZumbach91,KunzZumbach92}. In their remarkable 
work, Kunz and Zumbach~\cite{KunzZumbach92} concluded in 1992 in favor of 
such a topological transition scenario, essentially on the basis of
qualitative arguments (pairing of topological defects at low temperature where
they carry most of the energy in the system, sharp increase of the density
of defects and apparent discontinuity of the rotational
rigidity modulus at the transition, finite cusp in the specific heat, 
proliferation of unbinded defects at high temperature).  
Even though they performed a careful and sizeable study, 
they were unfortunately 
not able to decide conclusively between essential singularities or 
standard power laws - though their preference was for the first case - for 
the correlation length and the susceptibility 
when approaching the transition from the high temperature phase.
This is essentially due to the limited possibilities of computers ten years
ago, since the authors already took care about potential critical slowing
down problems as they adapted the Wolff cluster algorithm to the $RP^{n-1}$
model.

Ten years later, we want to address the same question of the nature of the
phase transition of the two-dimensional Lebwohl-Lasher model. Since we believe
that the conclusions of Kunz and Zumbach will hardly be improved, even with
more powerful facilities, it is necessary to reconsider the problem from a 
different point of view. We will thus assume the existence of a critical phase 
at low temperatures and then follow the line of the behaviour of the $XY$ 
model to predict consequences of the above mentioned assumption,
consequences which may
be compared easily to numerical results. Of course, the test 
must discriminate between different scenarios, starting
from the confirmation (and thus the characterization) of a topological
transition, a conventional continuous transition, a first-order one, or
no transition at all. The existence of a scale-invariant low temperature
critical phase is characteristic from the first situation. Such a system must
thus be conformally invariant at any temperature below the transition
$T_{KT}$ (we will abusively keep the terminology adapted to the $XY$ model),
so it becomes advantageous to deduce the functional expression of the 
correlation functions or density profiles in a restricted geometry adapted
to numerical simulations from a conformal mapping $w(z)$:
\be
\langle\bsigma_{w_1}\cdot\bsigma_{w_2}\rangle=|w'(z_1)|^{-x_\sigma}
|w'(z_2)|^{-x_\sigma}
\langle\bsigma_{z_1}\cdot\bsigma_{z_2}\rangle.
\label{covconf}
\ee
Here, $w$ labels the lattice sites in the transformed geometry (the one
where the computations are really performed), $z$ are the
corresponding points in the original one (usually the infinite plane where
$\langle\bsigma_{z_1}\cdot\bsigma_{z_2}\rangle\sim |z_1-z_2|^{-\eta_\sigma}$
takes the standard power-law expression), and $x_\sigma=\frac 12\eta_\sigma$
is the scaling dimension associated to the scaling field $\sigma$.
The interest of such an
approach lies in the full inclusion in the functional
expression of the changes due to shape effects. 
The most famous example is the exponential
decay of the correlation functions at criticality along a strip of finite 
width, unlike the algebraic decay in the infinite plane.
For simplicity reasons, it is even more convenient to work with density
profiles in a finite system with symmetry breaking fields along some surfaces
in order to induce a non vanishing local order parameter in the bulk.
In the case of a square lattice $\Lambda$ of size $L\times L$, with fixed
boundary conditions along the four edges $\partial\Lambda$, one 
expects~\cite{BurkhardtXue91,Berche02,Berche02b}
\begin{eqnarray}
m_{\Fbc}(w)&=&\langle P_2(\bsigma_w\cdot{\bf h}_{\partial\Lambda(w)})
\rangle_{\Fbc} 
\sim[\kappa(w)]^{-\frac 12\eta_\sigma}
\label{eq_m-kappa}\\
      \kappa(w)&=& {\rm Im}\left[{\rm sn}{\frac{2{\rm K}w}{L}} \right]\nonumber\\
      &\textstyle\times&\textstyle
        \left| \left(1-{\rm sn}^2{\scriptstyle\frac{2{\rm K}w}{L}}
        \right)  \left( 1-k^2{\rm sn}^2{\scriptstyle\frac{2{\rm K}w}{L}} \right)
\right|^{-\frac 12}
\end{eqnarray}
where $\Fbc$ specifies that the boundary conditions are fixed.
This expression easily follows from the expression of the order parameter 
profile decaying in the upper half-plane from a distant surface of spins 
constantly fixed in a given direction, 
$m(z)=\langle P_2(\bsigma_z\cdot{\bf h}_{\partial\Lambda(z)})\rangle_{\uhp} 
\sim  y^{-x_\sigma}$, and from the 
conformal transformation of the upper half-plane (\uhp) $z=x+iy$ ($0\le y<\infty$) 
inside a square $w=u+iv$ of size
$L\times L$ ($-L/2\le u\le L/2$, $0\le v\le L$)
with open boundary conditions
along the four edges, realized by a Schwarz-Christoffel 
transformation~\cite{LavrentievChabat}
\be
w(z)={L\over 2{\rm K}}{\rm F}(z,k),
\quad z={\rm sn}\left({2{\rm K}w\over L}\right).
\label{eq-SchChr}
\ee
Here, $F(z,k)$ is the elliptic integral of the 
first kind, 
${\rm sn}\ \! (2{\rm K}w/ L)$ 
the Jacobian elliptic sine, ${\rm K}=K(k)=F(1,k)$ the
complete elliptic integral of the first kind, 
and the modulus $k$ depends on the aspect ratio
of $\Lambda$.

Our strategy is now to fit numerical 
data of the order parameter profile against
expression~(\ref{eq_m-kappa}).
Like in the previous study of Kunz and 
Zumbach~\cite{KunzZumbach92,KunzZumbach91} the resort to a cluster 
update algorithm is necessary in order to prevent
the critical slowing down, all the spins of clusters (build through 
intermediate bond variables) being updated simultaneously. The algorithm
becomes particularly efficient if the percolation 
threshold of the bond process occurs at the transition 
temperature of the spin model, which 
ensures the updating of clusters of all sizes in a single
MC sweep. For $O(n)$ models, Ising variables are defined in the Wolff 
algorithm by the sign
of the projection of the spin variables along some random direction.
The bonds are then introduced through the Kasteleyn-Fortuin 
random graph representation~\cite{FortuinKasteleyn72}.
When one uses fixed boundary
conditions, a difficulty occurs and the Wolff algorithm should
become less efficient, since close to criticality the 
unique cluster will often
reach the boundary and no update is made in this case. 
This is circumvented by the following trick: even when the cluster 
reaches the fixed boundaries, 
it is updated - and so are the boundary spins -  
and the order parameter profile is then 
measured with respect to the
new direction  of the boundary
spins, $m_\Fbc(w)=\langle P_2(\bsigma_w\cdot\bsigma_{\partial\Lambda})
\rangle_{\Fbc}$.

Using this procedure, we studied systems of size $48\times 48$ up to $200\times 200$. For the
measurement of the order parameter profile,  
we discarded $10^6$ Wolff sweeps for thermalization, 
and the measurements were
performed on $10^6$ production sweeps. For reasons which are made obvious
below, the energy density required a better statistics and the
measurements were obtained over $16.10^6$ sweeps.

In order to underscore the existence of a line of marginal 
fixed points in the low temperature phase,
we first check qualitatively the expression of the energy-energy 
correlations from the behaviour of
the energy density profile. 
The energy density at site $w$ is for example 
defined as the average value of the energies
of the four links reaching $w$:
\be
\varepsilon_w=\frac {1}{2d}\sum_{\mu}
\left[P_2(\bsigma_{w-\mu}\cdot\bsigma_{w})+
P_2(\bsigma_w\cdot\bsigma_{w+\mu})\right].
\ee
The existence of a regular contribution in the 
energy density makes the calculation a bit more subtle than what presented
in equation~(\ref{eq_m-kappa}). 
This regular contribution $\langle\varepsilon_0(T)\rangle$ which 
depends on $T$ cancels after a suitable difference between
profiles obtained with different conditions at the boundaries (free 
($\fbc$) and fixed ($\Fbc$) boundary 
conditions). Although it makes the numerical computation longer in order
to reach some satisfactory accuracy, this makes possible to 
extract the singularity associated to the energy density:
\begin{eqnarray}
\langle\varepsilon_z\rangle_{\Fbc}&=&
\langle\varepsilon_0(T)\rangle+{\cal B}_{\Fbc}(T) y^{-\eta_\varepsilon(T)/2},\\
\langle\varepsilon_z\rangle_{\fbc}&=&
\langle\varepsilon_0(T)\rangle+{\cal B}_{\fbc}(T) y^{-\eta_\varepsilon(T)/2}.
\label{eq:E}
\end{eqnarray}
This is clearly illustrated in figure~\ref{fig1}1 where 
convergence towards the
same temperature-dependent constant $\langle\varepsilon_0(T)\rangle$ is shown,
with amplitudes of the singular 
terms having opposite signs, therefore 
a simple difference of the quantities measured in the square geometry,
\be
\Delta \varepsilon (w)=\langle\varepsilon_w\rangle_{\Fbc}-
\langle\varepsilon_w\rangle_{\fbc}
\sim \Delta{\cal B} \times [\kappa(w)]^{-\frac 12\eta_\varepsilon(T)}
\ee
leads to the value of the thermal scaling dimension $\eta_\varepsilon(T)$.
The right part in figure~\ref{fig1}1 presents a log-log plot of the 
difference $\Delta \varepsilon (w)$ {\it vs} $\kappa(w)$ at 
two temperatures below $T_{\rm KT}$ and one above which shows that the
functional expression used is no longer valid, as expected, in the 
paramagnetic phase. 
Due to the strong fluctuations, in the QLRO phase
the data scatter around straight lines which represent the slopes
$[\kappa(w)]^{-2}$. This figure, though not definitely conclusive,
confirms that the  exponent
of the decay of energy-energy correlations keeps a constant value
$\eta_\varepsilon(T)=4$ in the low-temperature phase of the $RP^2$ model,
confirming that like in the case 
of the $XY$ model, the temperature is a marginal field, responsible for the
existence of a critical line in the whole low-temperature phase. It thus
implies a thermal scaling exponent $x_\varepsilon=d-y_t=2$ which ensures
a vanishing RG eigenvalue $y_t=0$ (up to $T_{\rm KT}$ where it is consistent 
with an essential singularity of $\xi$
above the KT point, as suspected by Kunz and Zumbach~\cite{KunzZumbach92}). 
The energy-energy correlation function in the plane
should thus decay algebraically as
\be
\langle\varepsilon_{z_1}\varepsilon_{z_2}\rangle\sim
|z_1-z_2|^{-\eta_\varepsilon},\ee
with $\eta_\varepsilon(T)=2x_\varepsilon=4$ $\forall T<T_{\rm KT}$.

        \epsfysize=5.9cm
        \begin{center}
        \mbox{\epsfbox{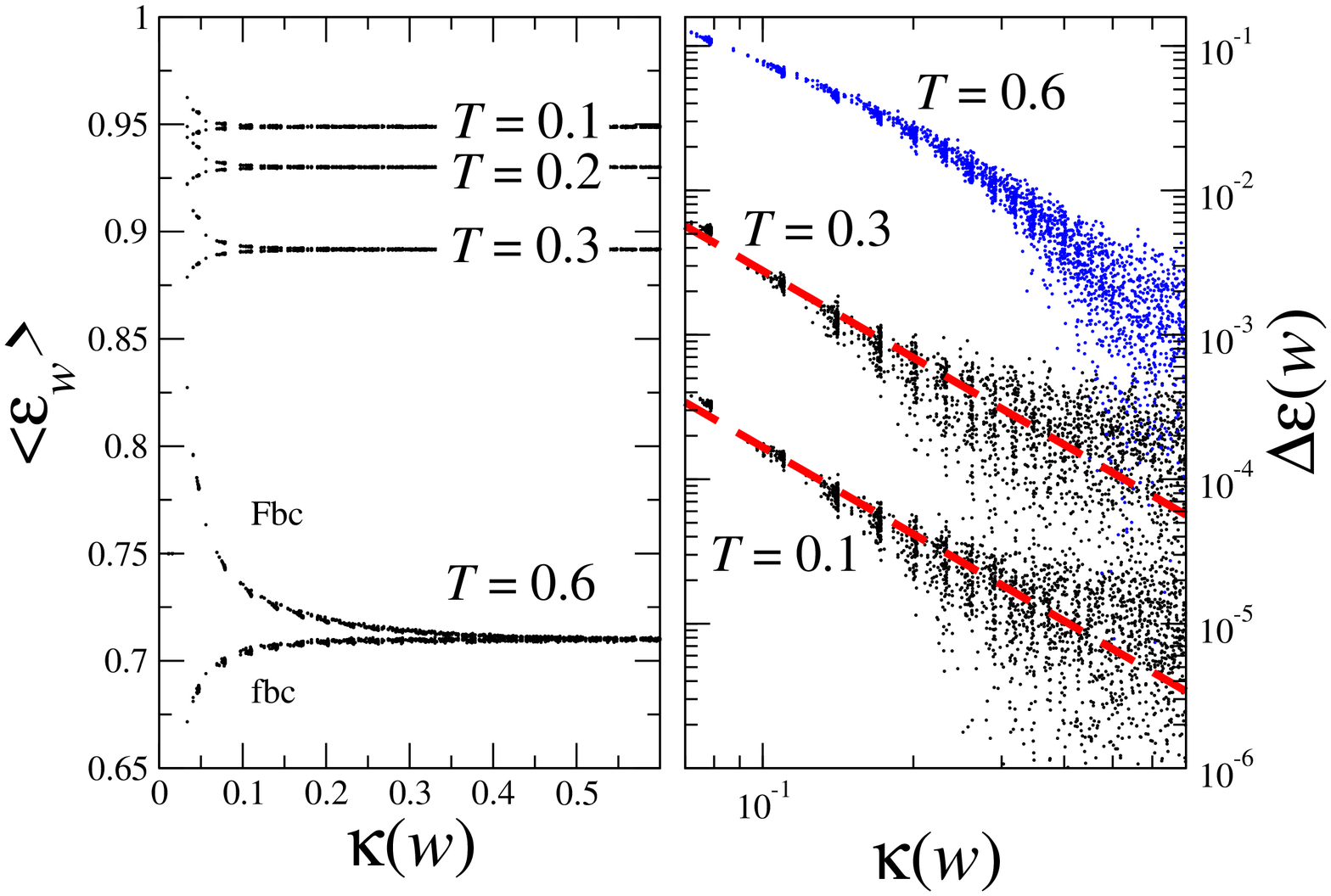}\qquad}
        \end{center}
        {{\small\noindent {\bf Fig. 1:} MC simulations of the $2d$ $RP^2$ 
            model inside a square of
        $100\times 100$ spins ($16.10^6$ MC sweeps after cancellation of 
        $10^6$ for thermalization). 
        Several temperatures below the transition temperature are shown, and one above ($T=0.6$).
        Left:  local energy density {\it vs} the 
        rescaled variable $\kappa(w)$.
        Right:  log-log plot of the 
        difference $\Delta \varepsilon(w)$.
        }}
        \label{fig1}  \vskip -0cm

From the existence of a marginal line, one may suspect that the other scaling
dimensions should continuously vary with the marginal field. 
The order parameter profile thus has to obey  equation~(\ref{eq_m-kappa}) 
in the whole
low temperature phase, but with an exponent $\eta_\sigma(T)$ which depends on
$T$. Equivalently, a log-log plot of $m_\Fbc(w)$ {\it vs} $\kappa(w)$ must
display straight lines with different slopes below $T_{\rm KT}$.
This is exactly what is observed in figure~\ref{fig2}2. Again, the curves
start to deviate from the straight line when the system enters the high 
temperature phase.

        \epsfysize=5.9cm
        \begin{center}
        \mbox{\epsfbox{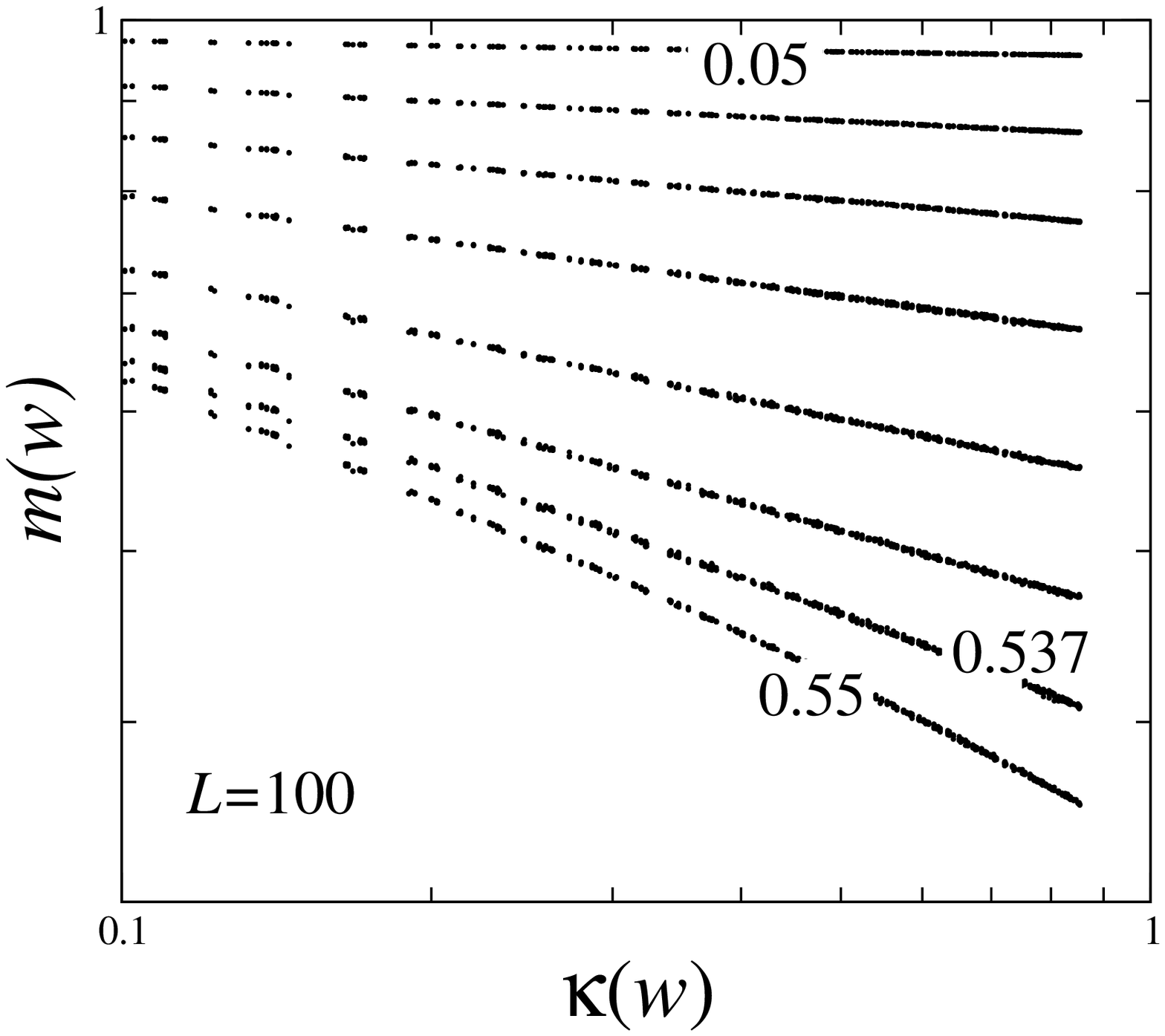}\qquad}
        \end{center}
        \noindent{{\small {\bf Fig. 2:} MC simulations of the $2d$ $RP^2$ model 
          inside a square of
        $100\times 100$ spins ($4.10^6$ MC sweeps after cancellation of $10^6$ for
        thermalization). The order parameter density is plotted
        against the rescaled
        distance at  different 
        temperatures below the KT
        transition temperature and slightly above. The numbers on the right give the value of
        $k_BT/ J $, and 0.537 (the estimate of Kunz and Zumbach~\cite{KunzZumbach92})
        appears to be overestimated.
        }}
        \label{fig2}  \vskip -0cm

        \epsfysize=6.6cm
        \begin{center}
        \mbox{\epsfbox{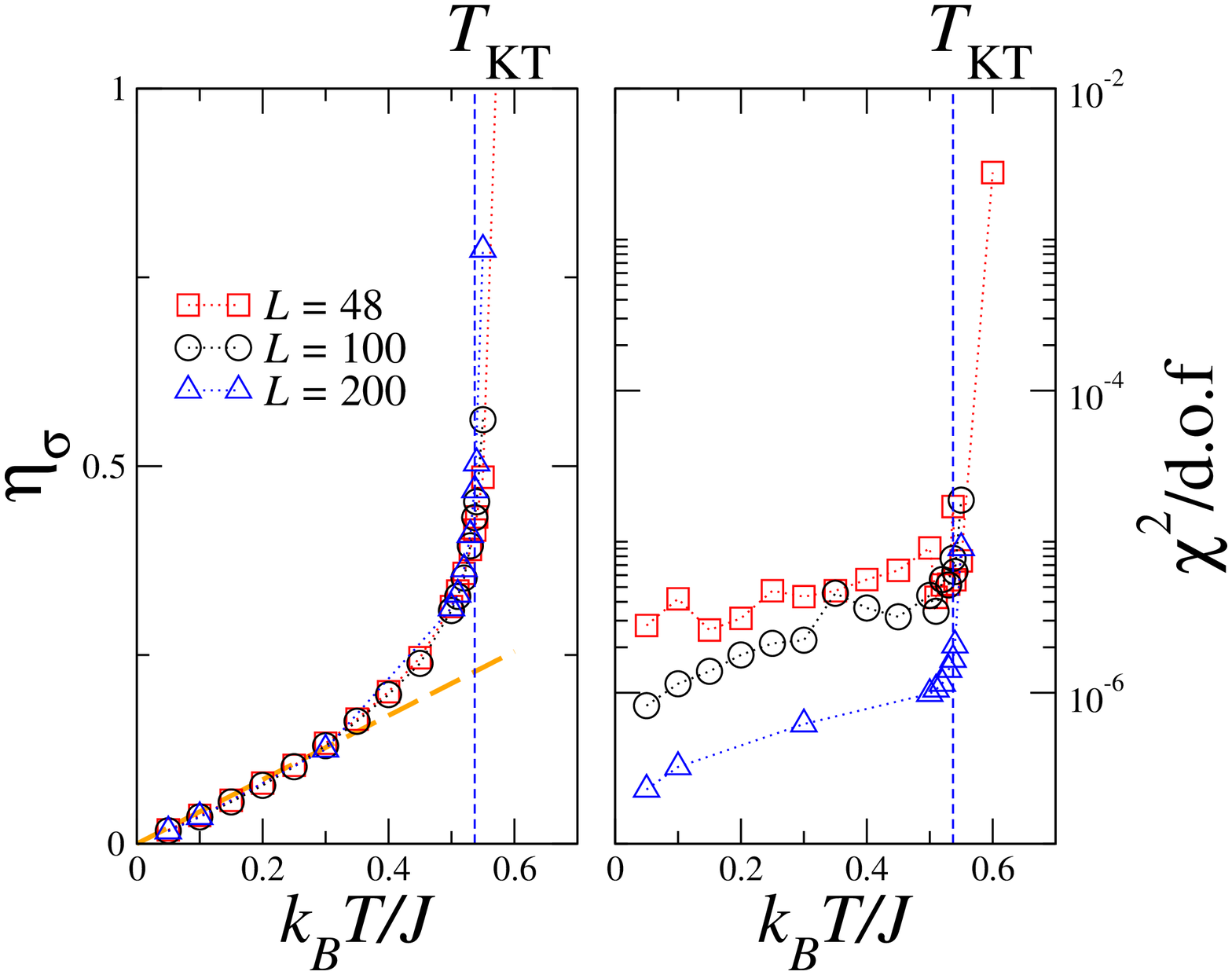}\qquad}
        \end{center}
        \noindent {{\small {\bf Fig. 3:} Left: Correlation function decay exponent of the $2d$ $RP^2$ 
          model as a function of the temperature. The vertical dashed line is the estimate of the
        transition temperature of Kunz and Zumbach~\cite{KunzZumbach92}. Right: value of the
      $\chi^2$ per degree of freedom resulting from the power law fits of the data in figure~\ref{fig2}2.
      The sharp increase is an indication of the location of the transition temperature (already below
      the estimate of Kunz and Zumbach~\cite{KunzZumbach92}).}}
        \label{fig3}  \vskip -0cm

The slopes measured in figure~\ref{fig2}2 are reported as a function of
the temperature in figure~\ref{fig3}3. 
        \epsfysize=5.9cm
        \begin{center}
        \mbox{\epsfbox{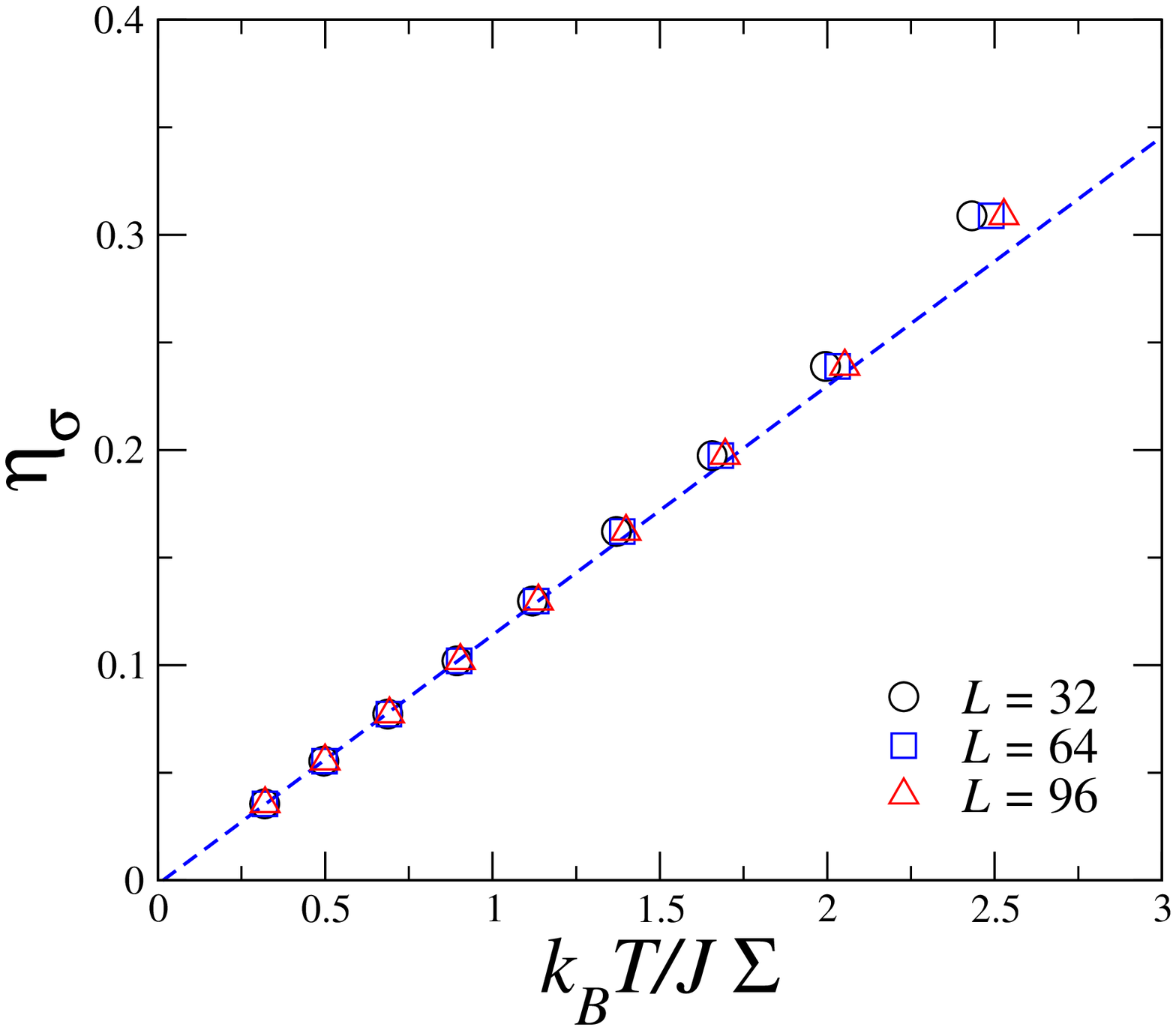}\qquad}
        \end{center}
        \noindent {{\small {\bf Fig. 4:}  Dependence of the rotational rigidity modulus with the 
            $\eta_\sigma$ exponent. }}
        \label{fig4}  \vskip -0cm

Another signature of this mechanism is the behaviour of the rotational rigidity modulus.
 This latter quantity generalizes the helicity 
modulus $\Upsilon$ in the $XY$ model, which measures the quadratic
response in the free energy of the system to a twist accross the sample. This is generalized
to the $RP^2$ model by measuring the change in free energy when a rotation of angle
$\phi$ around some axis in spin space is applied to the system: $F(\phi)-F(0)=\Sigma\phi^2
+O(\phi^4)$. This expression defines the rotational rigidity modulus $\Sigma$.
In the $XY$ model, there exists a universal relation between the helicity modulus and the
correlation function exponent, $\eta_\sigma=k_BT/ 2\pi J\Upsilon$~\cite{NelsonKosterlitz77}, which
appears as a consequence of the Kosterlitz recursion relations.
In the Lebwohl-Lasher model, the same type of behaviour is checked in figure~\ref{fig4}4 where
one observes a linear dependence (the larger the system size, the better the linear behaviour) 
of $\eta_\sigma$ with $T/\Sigma$, correctly fitted at low temperatures by
$\eta_\sigma=0.117\ k_BT/J \Sigma$.

To conclude, we mention that the most remarkable feature is a
complete analogy with the two-dimensional $XY$ model~\cite{Berche02} where the 
transition is mediated by the defects. The exponent
$\eta_\varepsilon$ keeps a constant value while $\eta_\sigma$, associated
to the order parameter, starts from zero at $T=0$ and increases
linearly with $T$ at low temperatures where a spin wave approximation
should capture the essentials of the behaviour of the system.
The influence of pairs of topological defects, which would appear in increasing 
number  when the temperature increases, is probably responsible for the
deviation from the spin wave approximation and of the sharper increase of
$\eta_\sigma$, and the transition is presumably completely governed
by unbinding of these defects, like in the Kosterlitz-Thouless scenario. The order parameter
exponent at the transition $k_BT_{\rm KT}/J\simeq 0.52$ 
takes a value $\eta_\sigma(T_{})\simeq 0.40(2)$.
The relation between rotational rigidity modulus and the exponent $\eta_\sigma(T)$
also seems  completely coherent with what happens in the $XY$ model, giving one more
evidence of the topological nature of the transition.

{\it Acknowledgement:} This work is supported by the french-venezuelian PCP
program  `Fluides p\'etroliers'. Support from the CINES under project c20020622309 is also
gratefully acknowledged.

\end{multicols}
\end{document}